# An expert system for diagnosing and treating heart disease

**Blake Fernandino, Moein Samak Bisheh**

Timely detection of illnesses is vital to prevent severe infections and ensure effective treatment, as it's always better to prevent diseases than to cure them. Sadly, many patients remain undiagnosed until their conditions worsen, resulting in high death rates. Expert systems offer a solution by automating early-stage diagnoses using a fuzzy rule-based approach. Our study gathered data from various sources, including hospitals, to develop an expert system aimed at identifying early signs of diseases, particularly heart conditions. The diagnostic process involves collecting and processing test results using the expert system, which categorizes disease risks and aids physicians in treatment decisions. By incorporating expert systems into clinical practice, we can improve the accuracy of disease detection and address challenges in patient management, particularly in areas with limited medical resources.

## 1. Introduction

Fuzzy logic emerges as a leading method for early disease detection through the analysis of patients' lab tests, facilitating swift and accurate diagnosis of heart conditions by physicians. A study by Pavate et al. (2019) outlined an online diabetes diagnosis system, focusing on understanding potential complications like blindness, kidney damage, heart attack, stroke, and other heart issues. Using a fuzzy-based model, their system achieved a notable 92.5% prediction accuracy. Patients could input symptoms online for precise recommendations based on the database.

Khairina et al. (2018) explored an automated disease diagnosis system using fuzzy rules, drawing on user knowledge and global databases. Similarly, Srivastava and Sharma (2019) shed light on heart disease phases due to sedentary lifestyles. Avci and Dogantekin (2016) focused on Parkinson's disease treatment with a genetic algorithm-based expert system. Santhanam and Ephzibah (2015) developed a hybrid genetic-fuzzy system for heart disease diagnosis, while Nahato et al. (2015) advocated for mathematical data models using data mining rules, employing BPNN for disease detection.

The overarching goal is to enhance existing expert systems' accuracy and efficiency using fuzzy logic, improving medical diagnoses from error-prone manual methods to precise fuzzy-based approximations, particularly for early disease identification and improved patient care.

## 2. Related work

The study conducted by Pavate et al. (2019) outlined an online diagnostic system tailored for identifying diabetes, with a key focus on understanding the potential complications associated with the disease's impact on the body. Their primary aim was to anticipate five major complications linked to diabetes: vision impairment, kidney damage, heart attack, stroke, and other cardiovascular issues. Through the use of a fuzzy-based model, their system achieved an impressive prediction accuracy of 92.5% via online access. Patients could input their symptoms on the web platform, enabling precise matching with the database and receiving customized recommendations based on their inputs.

In a separate inquiry, researchers explored the application of a multi-layered Mamdani fuzzy inference system (MI-MFIS) for diagnosing different stages of hepatitis B. This system, incorporating two input variables initially and seven in subsequent stages, evaluated liver health and hepatitis presence using data on enzyme vaccination collected by pathology departments.

Shiban and Meyer (2019) provided practical methods and clinical insights for therapy decision-making, emphasizing the pivotal role of randomized controlled trials (RCTs) in guiding treatment choices. Martinez et al. (2019) proposed a methodology involving data preprocessing, feature extraction, and feature reduction using Principal Component Analysis (PCA) and optimal classification, which proved effective in disease understanding.

Duţu, Mauris, and Bolon (2017) addressed issues surrounding fuzzy logic rule creation by proposing a precise and efficient rule-based expert system to mitigate these challenges. Khairina et al. (2018) explored an expert system for automated disease diagnosis, employing automatic fuzzy rules-based inference engines to streamline diagnosis and alleviate the burden on physicians.

Zhu et al. (2019) advanced disease detection with Enzyme-Linked Immunosorbent Assay (ELISA), offering a swift and efficient method for measuring glucose levels in humans. Rigla et al. (2018) delved into diabetes management through the collection of glucose and insulin pumping data, leveraging AI-driven systems to support doctors in decision-making processes.

Dankwa-Mullan et al. (2019) underscored the global prevalence of diabetes and the role of AI and cognitive computing in disease management, highlighting applications such as automated retinal screening and predictive risk stratification for patient self-management tools.

## 3. Materials and methods

The study introduces an expert system designed for precise medical diagnosis, particularly aimed at determining the minimum disease level and offering optimal solutions for various medical conditions, notably heart diseases. This system gathers input through direct patient inquiries, translating them into fuzzy rules based on blood symptoms and comparing them against a knowledge base processed by the inference engine. Upon receiving input, the system matches it with pre-established rules; if a match is found, the rule is activated; otherwise, the input may be disregarded. Activated rules facilitate the diagnosis of heart disease levels using existing knowledge base data.

Following diagnosis, the system advises physicians on the disease level based on the user-inputted symptoms. Subsequently, the physician prescribes suitable medication dosages, taking cues from the recommendations provided by the medical expert system.

The process involves several steps for implementing rules:
1. Prompting input by questioning the patient to obtain a precise set of data.
2. Comparing the provided input with the expert system's knowledge base using the inference engine.
3. Diagnosing the disease based on established rules, particularly for heart disease.
4. Selecting the appropriate disease level post-diagnosis.
5. Recommending the correct medication dosage for the patient.

The evaluation of this system employed various tools and programming languages, including Visual Studio 2013, MATLAB, and languages such as C/C++/C#. Additionally, medical equipment such as Electrocardiogram (ECG), Holter monitoring, Echocardiogram, and Angiography were utilized.

The proposed system's operational workflow is illustrated in Fig.1, depicting how it receives user input after diagnosing symptoms of given medical conditions and offers actionable recommendations for physicians. This investigation scrutinizes the expert system's efficacy in diagnosing and managing heart disease.

### 3.1 Input Parameters

#### 3.1.1 Cholesterol
Cholesterol, found in all human cells, plays a vital role in food digestion and the synthesis of vitamin D and hormones. It occurs naturally in the body and has a waxy, fatty appearance (Julia et al., 2022).

#### 3.1.1.1 Types of Cholesterol
Cholesterol comes in two forms:
a. Low-density lipoproteins (LDL)
b. High-density lipoproteins (HDL)
LDL, known as "bad" cholesterol, is harmful to health, whereas HDL, termed "good" cholesterol, is beneficial for the body.

#### 3.1.1.2 Functions of Cholesterol
Cholesterol serves four primary functions:
a. Constituting cell membranes
b. Assisting in bile acid production for digestion
c. Synthesizing vitamin D
d. Facilitating hormone production

#### 3.1.1.3 Causes of High Cholesterol
High cholesterol poses a significant risk for heart disease and potential heart attacks. Elevated cholesterol levels can lead to narrowed arteries, hindering blood flow. It often exhibits no symptoms, emphasizing the importance of regular testing and treatment to mitigate risk factors (Grundy & Vega, 1990).

#### 3.1.1.4 Cholesterol Levels
• Normal: 100-129 mg/dL
• High: 130-159 mg/dL
• Very high: 160-189 mg/dL

To lower cholesterol levels, individuals may adopt the following measures:
Following a heart-healthy diet
Engaging in regular exercise
Managing weight
Quitting smoking

*3.1.1.5 Cholesterol Treatment*
Various methods are available for treating high cholesterol levels, including:
• Lipid-lowering therapy
• Statin medication safety

*3.1.2 Blood Pressure*
The heart pumps blood to all body organs, exerting pressure on blood vessel walls. Blood pressure, measured in millimeters of mercury (mmHg), encompasses systolic and diastolic values (Mirmozaffari, 2019).

*3.1.2.1 Blood Pressure Ranges*
Normal blood pressure ranges between 90/60 mmHg and 120/80 mmHg. Values exceeding 140/90 mmHg indicate high blood pressure. Monitoring should occur throughout the day due to fluctuations influenced by stress or physical activity. Digital blood pressure monitors offer convenience but require proper positioning for accurate readings (Rohan & Venkadeshwaran, 2022).

*3.1.3 Chest Pain*
Also known as acute abdominal pain, severe abdominal pain warrants immediate attention as it may signal underlying heart issues. Possible causes include infection, inflammation, organ rupture, or blood flow obstruction (Kucia, Beltrame, & Keenan, 2022).

*3.1.3.1 Chest Pain Symptoms*

Low blood pressure
Progressive pain intensity
Abdominal swelling
Severe abdominal pain
Difficulty breathing
Sudden vomiting
Central abdomen pain
Profuse sweating
3.1.4 Blood Sugar
Blood sugar level refers to the amount of sugar in the blood, fluctuating based on diet, exercise, and glucose conversion. Elevated levels may indicate diabetes (Somogyi, 1945).

*3.1.5 Heart Rate*
Heart rate, measured in beats per minute, serves as an indicator of health and may reveal underlying health issues. It varies based on factors such as body weight, size, and environmental conditions (Mirmozaffari, 2019).

*3.1.6 Electrocardiogram*
An electrocardiogram (ECG) assesses heart function, detecting abnormalities and diagnosing heart attacks (Mincholé, Camps, Lyon, & Rodríguez, 2019).

*3.1.7 Glucose Meter*
A glucose meter measures blood glucose levels, aiding in diabetes management (Cao et al., 2022).

*3.1.8 Lipid Plus Device*
The lipid plus device assesses cholesterol levels, offering a cost-effective and user-friendly option for monitoring cholesterol levels (Bastianelli, Ledin, & Chen, 2017).

*3.1.9 Holter Monitor*
This device monitors heart rhythm over a 24-hour period, complementing ECG assessments (Mubarik & Iqbal, 2022).

*3.1.10 Blood Pressure Monitor*
Also known as a sphygmomanometer, this device measures blood pressure using an inflatable cuff and a measuring unit (Silva, Silva, & Guillo, 2020).

*3.2 Proposed System Flow*
The proposed expert system operates as follows:

Step 1: Input chest pain level and heart electrical signals.
Step 2: Input blood sugar level.
Step 3: Input heart rate.
Step 4: Input blood pressure.
Step 5: Input patient's gender and age.
Step 6: Input cholesterol value.
Step 7: Initiate diagnosis using the fuzzy-based expert system.
Step 8: Generate diagnosis report.
Step 9: Advise patient on appropriate medication dosage for heart disease treatment.
Step 10: Terminate system operation.

*3.2.1. Membership Functions for Fuzzy Rules*
Fuzzy rules have been crafted using the formula below for all membership functions of fuzzy sets:

Formula = $m^n$
  Here, "m" represents the membership function, and "n" signifies the input.

For ECG: There exist 3 membership functions (m) with 1 input (n).
Formula: 3^1 = 3
For Chest Pain: Similarly, there are 3 membership functions (m) with 1 input (n).
Formula: 3^1 = 3
For Blood Sugar: With 1 input (n), there are 4 membership functions (m).
Formula: 4^1 = 4
For Cholesterol: There are 3 membership functions (m) with 1 input (n).
Formula: 3^1 = 3
For Blood Pressure: Similar to cholesterol, there are 3 membership functions (m) with 1 input (n).
Formula: 3^1 = 3
For Heart Rate: With 1 input (n), there are 3 membership functions (m).
Formula: 3^1 = 3
For Age: Lastly, there are 4 membership functions (m) with 1 input (n).
Formula: 4^1 = 4

Hence, the total number of fuzzy rules has been calculated by multiplying the outputs of all membership functions, applying the aforementioned formula:

Total Rules = m1 * m2 * m3 * m4 * m5 * m6 * m7
= 3 * 3 * 4 * 3 * 3 * 3 * 4
= 3888

Therefore, a total of 3888 rules have been derived based on the given input parameters.

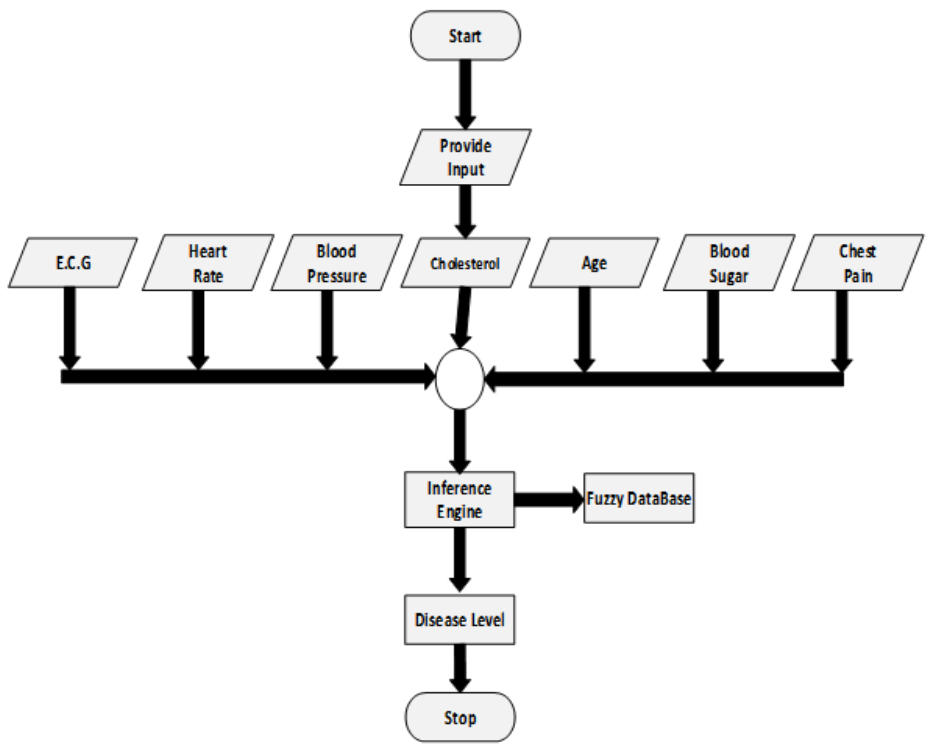

**Fig 1.** The System Flow Diagram

### 3.2.3. Fuzzy inputs and output
The fuzzy inputs and outputs are presented in Table 1.

**Table 1.**
The fuzzy input/output rule

| ECG | Chest Pain | Blood Sugar | Cholesterol | Blood Pressure | Age | Heart rate | Disease level |
|---|---|---|---|---|---|---|---|
| Medium | Typical Angina | Normal | Medium | High | Young | Medium | High |
| Normal | Normal | Medium | Medium | Medium | Young | Medium | Medium |
| Medium | Normal | Medium | Medium | Medium | Young | Medium | Medium |
| Normal | Normal | Normal | Normal | High | Young | Medium | Low |
| Medium | A Typical Angina | Normal | Medium | High | Aged | Medium | High |
| High | A Typical Angina | Normal | Medium | High | Aged | Medium | High |
| Medium | Normal | Medium | Medium | High | Aged | Medium | Medium |

### 3.2.4 Workflow of the Fuzzy-Based Expert System
1. Initialize the result to false.
2. Enter the ECG value.
3. Specify the level of chest pain.
4. Input the blood sugar level.
5. Provide the cholesterol value.
6. Enter the blood pressure level.
7. Input the age of the patient.
8. Input the heart rate of the patient.
9. Initiate the diagnosis process for the given parameters, including blood pressure, sugar level, and heart rate.
10. Determine the recommendation level.
11. Recommend the appropriate dosage to the patient.
12. Completion of setup 2.
13. Display true.
14. Exit.

### 3.3. Dataset
The dataset was obtained from THQ Yazman under the supervision of Dr. Bushra Zahid, the acting Medical Superintendent of THQ Yazman. Utilizing Visual Studio and MATLAB tools, the thesis research was implemented. With 3888 rules generated through fuzzy logic, the implementation phase commenced. An application was designed using the C# language to apply 100 out of the 3888 rules and obtain the desired output by providing values.

### 4. Results
Fuzzy logic was employed to generate all rules, which were then processed by MATLAB to yield results. Subsequently, the rules were implemented using Visual Studio Ultimate 2013 by Microsoft. The C# visual language facilitated suggestions and recommendations for both patients and doctors.

Data were collected from various sources such as hospitals, involving tests for patient age, gender, blood sugar, heart rate, sugar rate, and ECG. Upon completion of the testing process, diagnosis tests were conducted at the hospital by experts in heart disease. Crisp and fuzzy values were generated as inputs to the expert system. Following the acquisition of crisp input, the expert system commenced the fuzzification process. Subsequently, the expert system initiated the defuzzification process to convert fuzzy set values into a crisp format suitable for human interpretation.

After diagnosis by the expert system, physicians calculated values using fuzzy sets and provided outputs to determine the patient's heart disease. Upon completion of the diagnosis step, the expert system presented the output as the heart disease risk level categorized as Low, High, or Risky. Following the completion of the expert system's responsibilities, physicians determined treatment plans and recommended appropriate medication dosages based on the level determined by the expert system after diagnosis.

Simulation results demonstrated that this research achieved improved performance in determining proper heart disease risk levels, thus enhancing heart disease patient treatment. The simulation results were depicted to illustrate heart disease risk levels.

Fig.2 illustrates the ECG levels, depicting the output obtained after inputting ECG values and presenting the ECG level as normal, high, or very high.

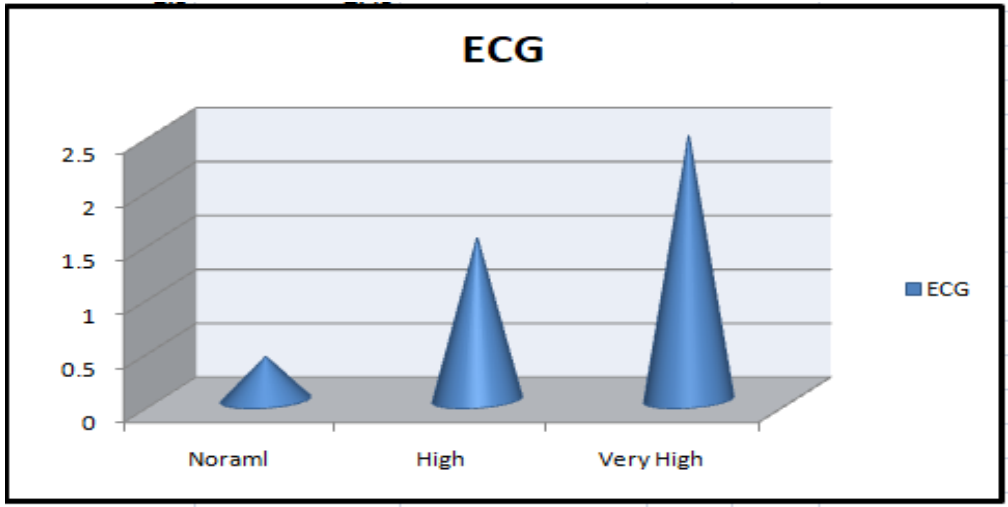

**Fig 2.** ECG levels

Fig.3 illustrates the levels of Chest Pain derived from the input provided by the doctor. After receiving the input, it categorizes the levels as Normal, High, and Very High. In Fig.4, the Blood Sugar levels are depicted by inputting data into the system, which subsequently generates results as output, categorized as Normal, High, and Very High.

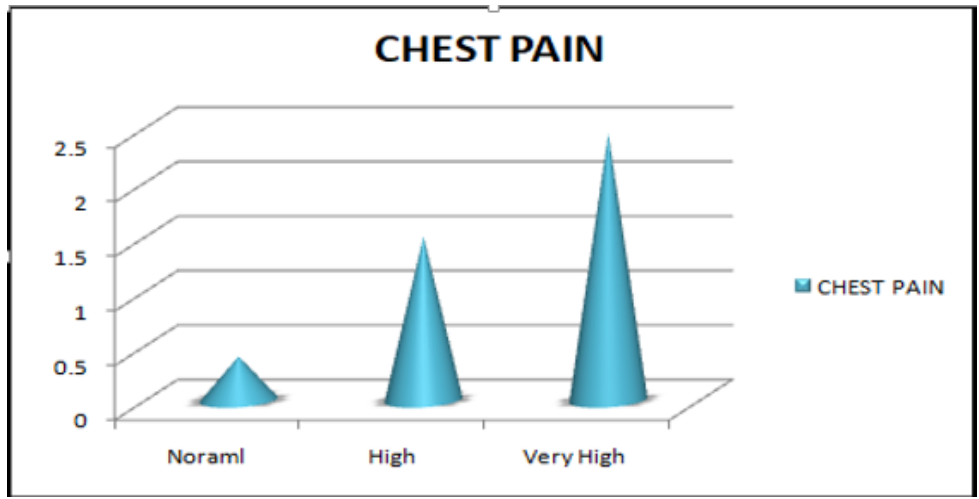

**Fig 3.** Chest Pain levels

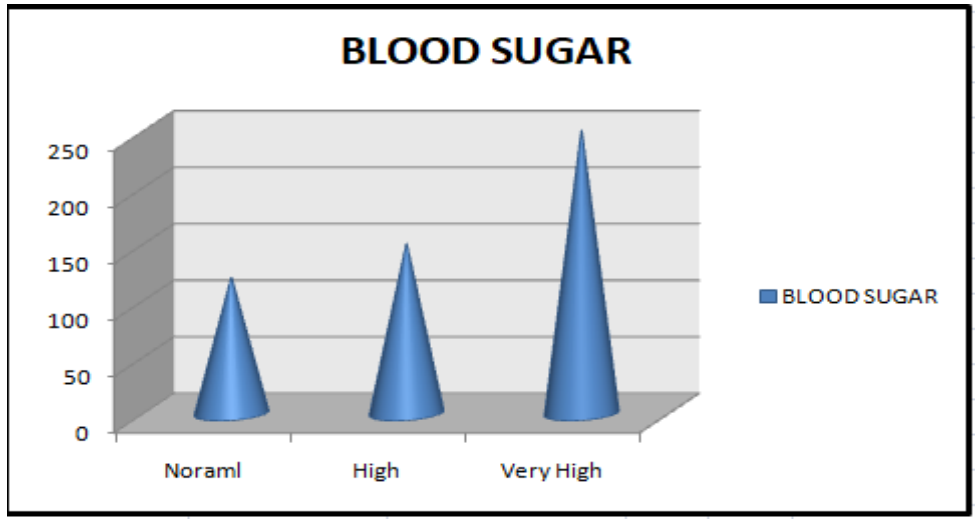

**Fig 4.** Blood sugar levels

Fig.5 displays the Cholesterol levels, presenting risk factors through its output, which delineates normal, high, or very high cholesterol levels. Fig.6 depicts the Blood Pressure levels, utilizing input from the physician doctor to determine the level of blood pressure.

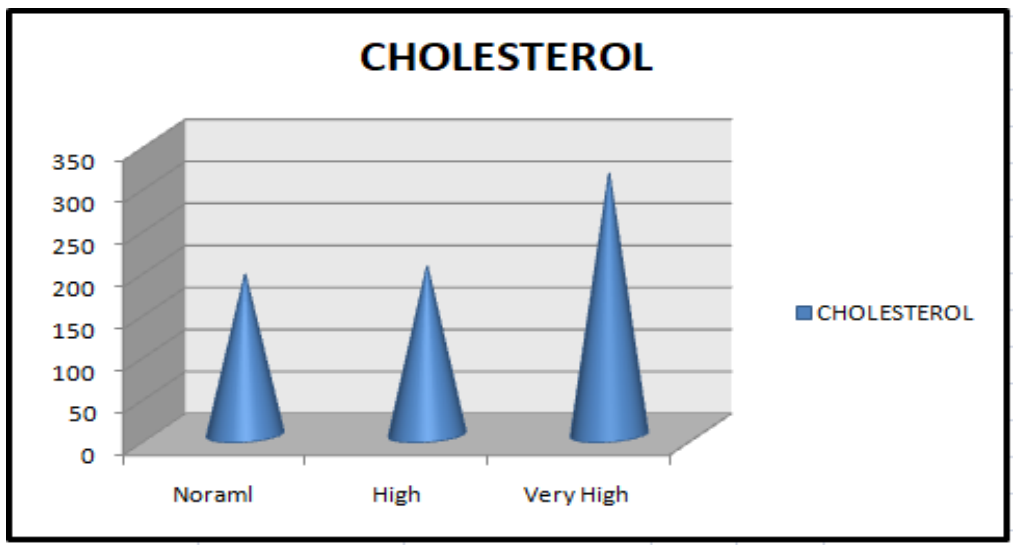

**Fig 5.** Cholesterol levels

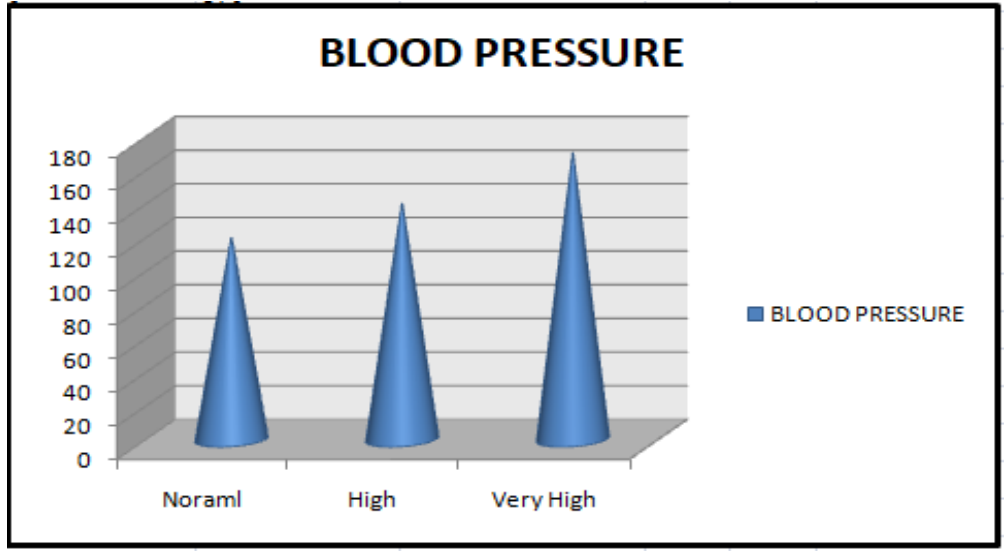

**Fig 6.** Blood Pressure levels

The Expert System for Heart Diseases Diagnosis and Treatment illustrates the age levels of the patient by receiving input from the user, as depicted in Fig.7. Users input the patient's age, and the system determines the disease level based on this factor. In Fig.8, the Heart Rate levels of the patient are displayed, utilizing results from various medical apparatus as input. The system then generates the heart rate level accordingly.

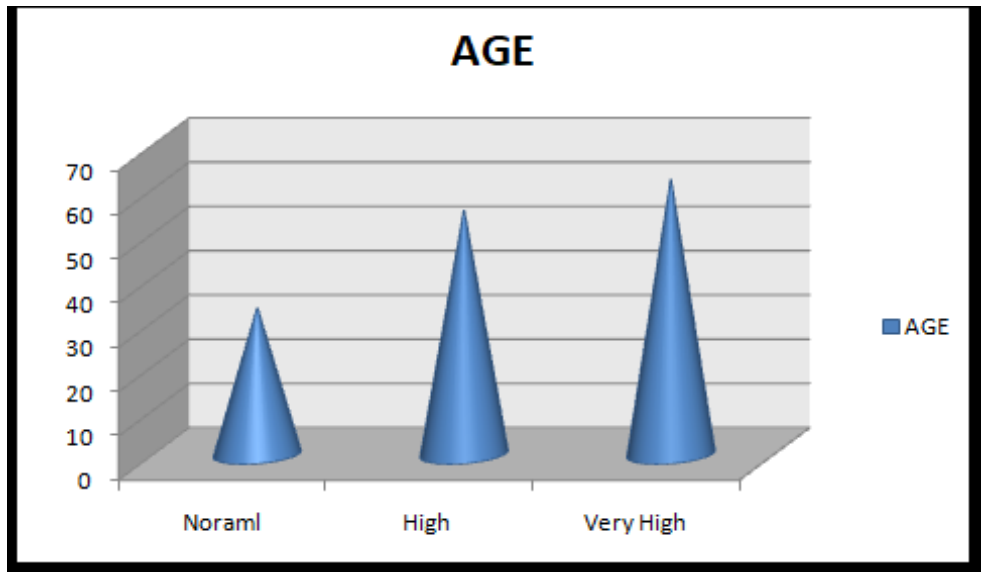

**Fig 7.** Levels of the patient's age

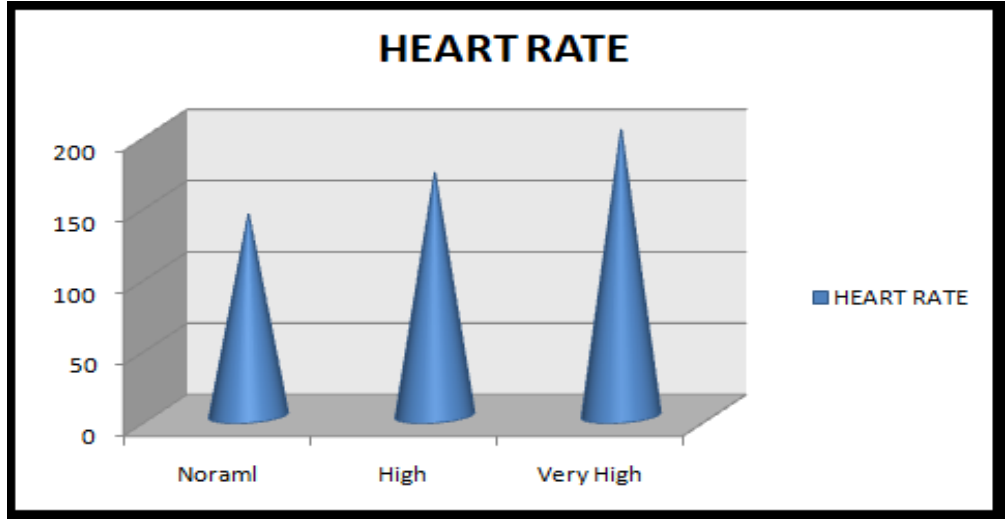

**Fig 8.** Patient's heart rate levels

The Expert System for Heart Diseases Diagnosis and Treatment displays the risk levels after receiving all inputs from the user. It then provides suggestions to the physician doctor regarding the level of heart disease, as illustrated in Fig.9).

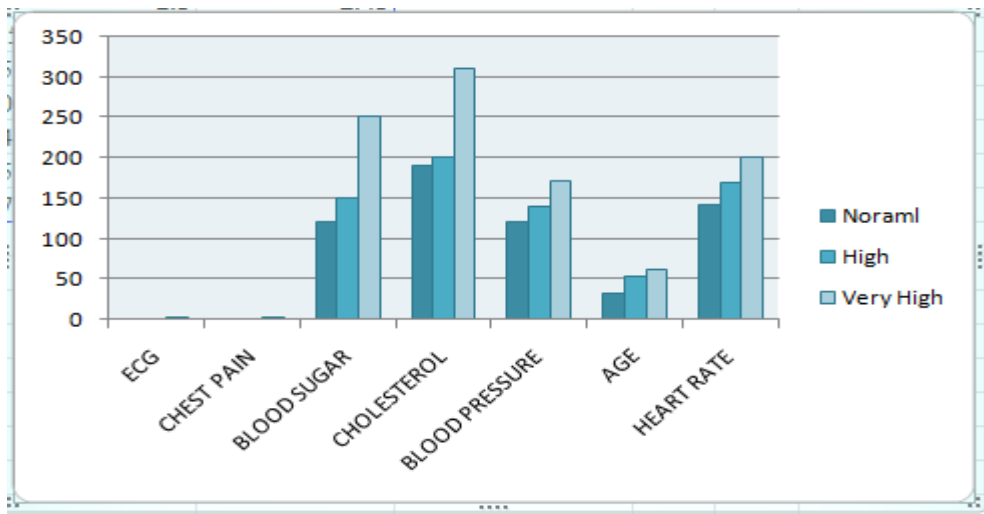

**Fig 9.** The levels for risk level output

### 4.1 Accuracy Level of Expert System

An expert opinion has been formulated based on multiple test results conducted on patients. The system achieves an average accuracy level of 95.5% after considering both the test results and expert opinion. This accuracy rate surpasses those of previously developed expert systems for various diseases. Table 2 presents the test results and expert opinion, along with detailing the accuracy level of the expert system.

Table 2
Test results and expert opinion

| Patient | ECG | Chest Pain | Blood Sugar | Cholesterol | Blood Pressure | Age | Heart rate | Human expert decision | expert system decision | Probability of correctness | Probability of errors |
|---|---|---|---|---|---|---|---|---|---|---|---|
| | mm/sec | ETT | mmol/L | mg/dL | mmHg | year | Bpm | | | | |
| 1 | 1.2 | 1.1 | 96 | 160 | 110 | 33 | 131 | Normal | Low | 0.95 | 0.05 |
| 2 | 1 | 1 | 102 | 130 | 115 | 30 | 136 | Normal | Low | 0.96 | 0.04 |
| 3 | 1.5 | 2 | 140 | 140 | 140 | 40 | 140 | Heart Disease found | Medium | 0.94 | 0.06 |
| 4 | 2.2 | 2.4 | 200 | 170 | 160 | 48 | 160 | Heart Disease found | High | 0.97 | 0.03 |
| 5 | 1.3 | 1.2 | 110 | 133 | 125 | 25 | 131 | Normal | Low | 0.95 | 0.05 |
| 6 | 1.4 | 1 | 109 | 117 | 126 | 23 | 127 | Normal | Low | 0.96 | 0.04 |
| 7 | 1.9 | 2.2 | 118 | 170 | 149 | 36 | 137 | Heart Disease found | Medium | 0.94 | 0.06 |
| 8 | 2.3 | 2.4 | 190 | 160 | 150 | 38 | 180 | Heart Disease found | High | 0.97 | 0.03 |
| 9 | 1.1 | 1.1 | 112 | 120 | 131 | 29 | 128 | Normal | Low | 0.95 | 0.05 |
| 10 | 1.3 | 1 | 115 | 109 | 121 | 35 | 119 | Normal | Low | 0.96 | 0.04 |

## 4. Conclusion

In the proposed research, all input is sourced directly from the patients and subsequently evaluated by expert doctors at the THQ Tehsil level hospital. The verification of all test results adds to the effectiveness of the system, rendering it more reliable and beneficial for both doctors and patients alike. This approach aids in alleviating the congestion at hospitals and minimizes the need for continuous patient monitoring. The implementation of a fuzzy expert system proves to be highly advantageous for diagnosing heart disease patients, offering valuable support to both patients and doctors. Fuzzy member functions and rules are developed using fuzzy logic, resulting in the creation of 3888 rules, albeit only 97 rules were ultimately utilized. The results were then implemented through a specially designed C# application tailored for the Expert system for heart disease diagnosis and treatment.